\documentclass[superscriptaddress,showpacs,preprintnumbers,amsmath,amssymb,
nofootinbib,twocolumn,aps,prd,10pt]{revtex4-1}
\usepackage{amssymb,amsmath}
\usepackage{epsfig}
\usepackage{xcolor}
\usepackage[utf8]{inputenc}
\usepackage{stmaryrd}
\usepackage{mathrsfs}
\usepackage{mathalfa}
\usepackage[normalem]{ulem}

\newcommand{\be}{\begin{equation}}
\newcommand{\ee}{\end{equation}}
\newcommand{\ba}{\begin{eqnarray}}
\newcommand{\ea}{\end{eqnarray}}
\newcommand{\beg}{\begin{gather*}}
\newcommand{\eng}{\end{gather*}}
\newcommand{\hh}{,\hspace{0.5cm}}
\newcommand{\hhh}{,\hspace{0.2cm}}

\newcommand{\lap}{\triangle}
\newcommand{\n}[1]{\label{#1}}

\newcommand{\ts}[1]{{\boldsymbol{#1}}}

\def\XXint#1#2#3{{\setbox0=\hbox{$#1{#2#3}{\int}$ }
\vcenter{\hbox{$#2#3$ }}\kern-.6\wd0}}

\usepackage[makeroom]{cancel}
\usepackage[caption=false]{subfig}
\usepackage[colorlinks=true,
            citecolor=green,
            linkcolor=red,
            filecolor=cyan,
            urlcolor=magenta,
            backref=false]{hyperref}

\newcommand{\dd}{\mbox{d}}

\begin{document}

\title{Ultrarelativistic spinning objects in non-local ghost-free gravity}

\author{Jens Boos}
\email{boos@ualberta.ca}
\affiliation{Theoretical Physics Institute, University of Alberta, Edmonton, Alberta, Canada T6G 2E1}
\author{Jose Pinedo Soto}
\email{pinedoso@ualberta.ca}
\affiliation{Theoretical Physics Institute, University of Alberta, Edmonton, Alberta, Canada T6G 2E1}
\author{Valeri P. Frolov}
\email{vfrolov@ualberta.ca}
\affiliation{Theoretical Physics Institute, University of Alberta, Edmonton, Alberta, Canada T6G 2E1}

\date{\today}

\begin{abstract}
We study the gravitational field of ultrarelativistic spinning objects (gyratons) in a modified gravity theory with higher derivatives. In particular, we focus on a special class of such theories with an infinite number of derivatives known as ``ghost-free gravity'' that include a non-local form factor such as $\exp(-\Box\ell^2)$, where $\ell$ is the scale of non-locality. First, we obtain solutions of the linearized ghost-free equations for stationary spinning objects. To obtain gyraton solutions we boost these metrics and take their Penrose limit. This approach allows us to perform calculations for any number of spacetime dimensions. All solutions are regular at the gyraton axis. In four dimensions, when the scale non-locality $\ell$ tends to zero, the obtained gyraton solutions correctly reproduce the Aichelburg--Sexl metric and its generalization to spinning sources found earlier by Bonnor. We also study the properties of the obtained four-dimensional and higher-dimensional ghost-free gyraton metrics and briefly discuss their possible applications.
\end{abstract}


\maketitle

\section{Introduction}
The study of the gravitational field of ultrarelativistic particles and beams of light is a very old subject. The first solution describing the gravitational field of beam of light (``pencil'') was found by Tolman, Ehrenfest and Podolski in 1931 \cite{PhysRev.37.602}. These authors used a linear approximation of the Einstein equations. One of their main conclusions was that the gravitational force acting on a massless particle moving in the same direction as the beam of light vanishes. Later, Bonnor \cite{Bonnor1969} presented a solution for the gravitational field produced by a cylindrical beam of a null fluid. This model can be interpreted as a description of a high frequency light beam in the geometric optics approximation when diffraction effects are neglected.\footnote{More recent studies of light beams beyond the geometric optics approximation can be found in \cite{Schneiter_2018} and references therein.} The gravitational field of a spinning pencil of light was obtained by Bonnor in 1970 \cite{Bonnor:1970sb}, see also Refs.~\cite{doi:10.1063/1.528380,PhysRevD.96.104053}. Higher-dimensional solutions describing the gravitational field of spinning ultrarelativistic objects and light beams were obtained in \cite{Frolov:2005in,Frolov:2005zq}. The latter work introduced the name ``gyraton'' for such spinning ultrarelativistic objects, which is now used in the literature quite frequently. There exist different generalizations of standard gyraton solutions, such as solutions for charged gyratons \cite{Frolov:2005ja}, gyratons in asymptotically AdS spacetimes \cite{Frolov:2005ww}, in a generalized Melvin universe with cosmological constant \cite{Kadlecova:2016irj}, and string gyratons in supergravity \cite{Frolov:2006va}. Gyraton solutions of the Einstein equations belong to the wide class of so-called Kundt metrics \cite{Stephani:2003tm}. A comprehensive discussion of gyratons in the Robinson--Trautman and Kundt classes of metrics can be found in \cite{Kadlecova:2009qu,Krtous:2012qa,Podolsky:2014lpa,Podolsky:2018oov}.

There is another problem that has been widely discussed in the literature and which is closely related to gyratons. In 1970, Aichelburg and Sexl \cite{Aichelburg:1970dh} constructed a metric of a massive ultrarelativistic particle. In its rest frame, the gravitational field of such a particle of mass $m$ is described by the Schwarschild metric. In order to obtain the metric when this particle moves with a very high velocity they applied a boost transformation and considered the limit wherein the velocity of the object tends to the speed of light, and hence the Lorentz factor $\gamma$ diverges. They demonstrated that keeping the value of the energy $E=\gamma m$ fixed yields a limiting metric which is now called the Aichelburg--Sexl solution. For this solution the gravitational field of a particle is localized at the null plane tangent to the null vector of the particle's four-velocity. Later, Penrose \cite{Penrose1976} demonstrated that this is a generic property of any metric that is boosted to the speed of light, provided the corresponding energy is kept fixed, and this special limiting case has hence been dubbed ``Penrose limit.'' Aichelburg--Sexl-type metrics have been widely used for the study of the gravitational interaction of two ultrarelativistic particles as well as black hole production via their collision. The area of the apparent horizon in this process just before the moment of collision was calculated in \cite{Eardley:2002re} and has been widely used for estimating black hole formation cross sections in the collision of ultrarelativistic particles (see e.g.~\cite{Yoshino:2002tx,Giddings:2004xy,Yoshino:2005hi,Yoshino:2006dp,Yoshino:2007ph} and references therein).

Since in the Penrose limit the initial mass $m$ of the particle tends to zero, one can obtain the Aichelburg--Sexl metric by starting with a linearized, weak-field gravity solution for a point-like particle. By considering a superposition of such solutions it is easy to construct the gravitational field of extended objects in linearized gravity. In particular, one may consider first a line distribution of mass, and then boost the solution. Due to the Lorentz contraction in the direction of motion the visible size of the body in this direction shrinks. This means that in order to obtain a solution for the ultrarelativistic case featuring a finite energy distribution profile one needs not only to take the Penrose limit keeping $\gamma m$ constant, but also simultaneously keep the parameter $L/\gamma$ fixed, where $L$ is the size of the object in the direction of motion. Such a procedure can be applied to a spinning object provided the rotation takes place within the plane orthogonal to the direction of motion. One can show that in such a procedure one reconstructs the gravitational field of a gyraton. This method is described in details in chapter 5 of the book \cite{Frolov:1418196}.

The goal of this paper is to construct gyraton-like solutions in so-called ``ghost-free'' gravity. This is an important special class of modified gravity theories that introduces non-locality by means of non-local form factors of the type $\exp[(-\Box\ell^2)^N]$. This modification becomes relevant only at small scales comparable to $\ell$, and hence this type of theories can be considered an ultrviolet (UV) modification of gravity. To that end, the main motivation for this study is that a small scale modification of gravity might become important for the process of mini black hole formation in the collision of ultrarelativistic particles. For example, it was shown that if the Einstein--Hilbert action is modified by the inclusion of higher-derivative as well as infinite-derivative terms, there exists a mass gap for black hole formation \cite{Frolov:2015bta,Frolov:2015bia,Frolov:2015usa,Giacchini:2018gxp}.

While non-locality has been explored for quite some time \cite{Yukawa:1949a,Yukawa:1949b,Yukawa:1950a,Yukawa:1950b,Efimov1967,Efimov:1972wj,Efimov:1976nu,Efimov:18,Efimov:19}, the particular class studied here is motivated from string theory \cite{Frampton:1988kr,Tseytlin:1995uq,Tomboulis:1997gg,Biswas:2005qr} as well as non-commutative geometry \cite{Spallucci:2006zj}. These non-local theories of gravity have appealing UV properties \cite{Biswas:2010yx,Modesto:2011kw} and are under active investigation. It has been demonstrated that in the weak-field regime this class of theories regularizes the gravitational field of point-like sources \cite{Edholm:2016hbt,Buoninfante:2018rlq,Buoninfante:2018stt,Giacchini:2018wlf} as well as thin brane-like extended objects \cite{Boos:2018bxf,Boos:2020kgj,Kolar:2020bpo}. For results in the strong-field regime in connection with black holes we refer to \cite{Conroy:2015wfa,Li:2015bqa,Calcagni:2017sov,Koshelev:2018hpt,Boos:2019vcz} and references therein; for cosmological applications see \cite{Biswas:2010zk,Calcagni:2013vra}. Non-local infinite-derivative form factors have also been explored in quantum theory \cite{Boos:2018kir,Buoninfante:2019teo} as well as quantum field theory \cite{Shapiro:2015uxa,Frolov:2016xhq,Modesto:2017hzl,Asorey:2018wot,Calcagni:2018gke,Buoninfante:2018mre,Boos:2019fbu}.

This paper is organized as follows: we begin by discussing the solutions of the modified gravity equations in the weak-field approximation. In Sec.~II we consider a wide class of theories for which the linearized action is quadratic in curvature and contains an arbitrary number of derivatives. General analysis of such theories shows that their action can be rewritten in a form which contains two scalar functions of the d'Alembert operator, where an additional requirement of the absence of scalar modes establishes a relation between these functions \cite{Biswas:2011ar}. We obtain a general solution of the field equations for a stationary distribution of spinning matter in four and higher dimensions, paying  special attention to extended pencil-type distribution of spinning matter. In Sec.~III we apply the boost transformation to these pencil-like distributions of matter choosing the velocity being directed along the pencil axis. After this, we obtain the Penrose limit for the boosted metrics and find the gravitational field of ultrarelativistic extended spinning objects (gyratons) in four and higher dimensions. The explicit form of these metrics and their properties are discussed in Sec.~IV. We summarize our findings and mention possible future applications of these solutions in Sec.~V.

\section{Spinning objects in the weak-field approximation of infinite-derivative gravity}
\label{sec:2}

\subsection{Linearized equations}

Our goal is to obtain the metric for an ultrarelativistic spinning object (gyraton) in infinite-derivative gravity. The method of solving the problem is the following. First one finds a solution of the gravitational field equations in the object's rest frame. Then one transforms this solution to a new reference frame moving with a constant velocity $v$ with respect to the original one. Finally, one takes the limit $v\to 1$ while keeping the energy $E=m\gamma=m/\sqrt{1-v^2}$ fixed. In this \emph{Penrose limit} the original mass $m$ of the object effectively approaches zero, which implies that in order to obtain the corresponding gyraton metric one may start with a solution with very small mass $m$. One can expect that higher-order curvature corrections---which are proportional to second and higher orders in the mass $m$---are therefore small. In this section we will discuss the field equations of linearized infinite-derivative gravity and present their solutions for an extended, slowly spinning object of small mass.

We denote by $X^{\mu}=(t,x^{\alpha})$ Cartesian coordinates in $(d+1)$-dimensional Minkowski spacetime and use indices $\alpha, \beta, \ldots =1,2,\ldots ,d$ from the beginning of the Greek alphabet to label spatial coordinates. The Minkowski metric in the $D=d+1$ dimensional spacetime is
\begin{align}
\dd s_0^2=\eta_{\mu\nu}\dd X^{\mu} \dd X^{\nu} = -\dd t^2 + \delta_{\alpha\beta}\dd x^{\alpha} \dd x^{\beta}\, ,
\end{align}
where $\delta{}_{\alpha\beta}$ denotes the flat $d$-dimensional metric. We denote by $h{}_{\mu\nu}$ a small deviation of the metric from the flat background,
\begin{align}
g{}_{\mu\nu} = \eta{}_{\mu\nu} + h{}_{\mu\nu} \, .
\end{align}
One can show that the most general linearized action in a Lorentz invariant theory with an arbitrary number of derivatives and quadratic in the perturbation $h_{\mu\nu}$ can be written in the form \cite{Biswas:2011ar}
\begin{align}
S &= \frac{1}{2\kappa} \int \dd^D x \Big( \frac12 h^{\mu\nu}\,a(\Box)\Box\,h_{\mu\nu}-h^{\mu\nu}\,a(\Box)\partial_{\mu}\partial_{\alpha}\,h^{\alpha}{}_{\nu} \nonumber \\
&\hspace{75pt} + h^{\mu\nu}\, c(\Box)\partial_{\mu}\partial_{\nu} h - \frac12 h\,c(\Box)\Box h \label{eq:action} \\
&\hspace{75pt} + \frac12 h^{\mu\nu}\,\frac{a(\Box)-c(\Box)}{\Box}\partial_{\mu}\partial_{\nu}\partial_{\alpha}\partial_{\beta}\,h^{\alpha\beta}\Big) \, , \nonumber
\end{align}
where $\Box$ is the d'Alembert operator of Minkowski space, $\Box = \eta{}^{\mu\nu}\partial_\mu\partial_\nu$.
The functions $a(\Box)$ and $c(\Box)$ can be chosen freely to parametrize different Lorentz-invariant modifications of gravity, subject only to the constraint
\begin{align}
\label{eq:constraint}
a(0) = c(0) = 1
\end{align}
which guarantees the proper Newtonian limit; see also the related discussions in Refs.~\cite{Biswas:2011ar,Buoninfante:2018mre}. In the case of $a(\Box) = c(\Box) = 1$ one recovers the Fierz--Pauli action and linearized General Relativity.

The field equations corresponding to the action \eqref{eq:action} are
\begin{align}
\begin{split}
\label{EQN}
&\hspace{11pt}a(\Box)\big[\Box\,h_{\mu\nu}-
\partial_{\sigma}\big(\partial_{\nu}\,h_{\mu}{}^{\sigma} +\partial_{\mu}h_{\nu}{}^{\sigma}\big)\big]\\
&+c(\Box)\big[\eta_{\mu\nu}\big(\partial_{\rho}\partial_{\sigma}h^{\rho\sigma}-\Box h\big)+\partial_{\mu}\partial_{\nu}h \big]\\
&+{a(\Box)-c(\Box)\over\Box}\partial_{\mu}\partial_{\nu}\partial_{\rho}\partial_{\sigma}h^{\rho\sigma}=-2\kappa  T_{\mu\nu} \, ,
\end{split}
\end{align}
where $T{}_{\mu\nu}$ is the energy-momentum tensor of matter, and $h = \eta{}^{\alpha\beta}h{}_{\alpha\beta}$ denotes the trace of $h{}_{\mu\nu}$. From now on we shall restrict ourselves to the case of
\begin{align}
c(\Box) = a(\Box) \, .
\end{align}
This condition guarantees that no extra scalar modes are present in the theory \cite{Biswas:2011ar}. We denote
\begin{align}
\hat{h}_{\mu\nu} = h{}_{\mu\nu} - \frac12 h \eta{}_{\mu\nu} \, .
\end{align}
The inverse transformation is
\begin{align}\n{inv}
h{}_{\mu\nu} = \hat{h}{}_{\mu\nu} - \frac{1}{d-1} \hat{h} \eta{}_{\mu\nu} \, .
\end{align}
We also impose the gauge conditions $\partial{}_\mu \hat{h}{}^{\mu\nu} = 0$. Then, Eq.~\eqref{EQN} simplifies greatly and takes the form
\begin{align}
\label{eq:eom}
a(\Box) \Box \hat{h}{}_{\mu\nu} &= -2\kappa T_{\mu\nu} \, .
\end{align}
The conservation law $\partial{}_\mu T^{\mu\nu}=0$ implies that the imposed gauge conditions are consistent.

\subsection{Stationary solutions for extended sources}
We assume that $T_{\mu\nu}$ does not depend on time. For a stationary metric generated by such a stress-energy tensor the $\Box$-operator reduces to the $d$-dimensional Laplace operator $\lap=\delta^{\alpha\beta}\partial_{\alpha}\partial_{\beta}$. We denote
\begin{align}
\mathcal{D}=a(\lap)\lap \, .
\end{align}
Then, we can solve the field equations \eqref{eq:eom} by using the static Green function
\begin{align}
\label{eq:gf-def}
\mathcal{D}\mathcal{G}_d(\ts{x},\ts{x'}) = -\delta{}^{(d)}(\ts{x}-\ts{x'}) \, .
\end{align}
The solution then takes the form
\begin{align}
\label{hhat}
\hat{h}{}_{\mu\nu}(\ts{x}) = 2\kappa \int \dd^d y \, \mathcal{G}_d(\ts{x}-\ts{y}) T{}_{\mu\nu}(\ts{y}) \, .
\end{align}
The expression for the perturbation of the metric $h_{\mu\nu}$ can be found from \eqref{hhat} by using relation \eqref{inv}.
For the stress-energy tensor \eqref{a1} given in the Appendix one has
\begin{align}
T_{\mu\nu}=\rho(\ts{x})\delta^t_{\mu}\delta^t_{\nu}+\delta^t_{(\mu}\delta^{\alpha}_{\nu)}
{\partial \over \partial x^ \beta} j_\alpha{}^\beta(\ts{x})\, .
\end{align}
A solution $h_{\mu\nu}$ of the field equations \eqref{eq:eom} for this source can be written as follows:
\begin{align}
\ts{h}&=h_{\mu\nu}\dd X^{\mu} \dd X^{\nu}\, , \label{solh} \\
\ts{h} & =\phi \left(\dd t^2+{1\over d-2} \delta_{\alpha\beta}\dd x^{\alpha} \dd x^{\beta} \right)
+2 A_{\alpha}\dd x^{\alpha}\dd t \, , \nonumber \\
{\phi}(\ts{x}) &= 2\kappa \frac{d-2}{d-1} \int \dd^d y \, {\rho}(\ts{y}) \mathcal{G}_d(\ts{x}-\ts{y}) \, , \label{solPhi} \\
{A}_{\alpha}(\ts{x}) &= \kappa \int \dd^d y \, {j}{}_{\alpha}{{}^{\beta}(\ts{y}) \frac{\partial \mathcal{G}_d(\ts{x}-\ts{y})}{\partial x{}^{\beta}}} \, . \label{solA}
\end{align}

Due to the translational symmetry of Eq.~\eqref{eq:gf-def}, the Green function $\mathcal{G}_d(\ts{x},\ts{x'})$ is a function of $\ts{x}-\ts{x'}$, while due to the spherical symmetry it depends on the radius variable $r=|\ts{x}-\ts{x'}|$
alone. Thus one has\footnote{In order to keep the notation somewhat manageable we shall use the same symbol for the Green function with vectorial arguments $\mathcal{G}_d(\ts{x})$ and with the scalar radius argument $\mathcal{G}_d(r)$.}
\begin{align}
\mathcal{G}_d(\ts{x}-\ts{x'})=\mathcal{G}_d(r)\, .
\end{align}
As has been shown previously \cite{Boos:2018bxf}, the Green function  in $d+2$ spatial dimensions is related to the Green function in $d$ spatial dimensions via the recursion formulas
\begin{align}
\label{eq:gf-rec-1}
\mathcal{G}_d(r) &= -2\pi\int \dd\tilde{r} \, \tilde{r} \, \mathcal{G}_{d+2}(\tilde{r}) \, , \\
\label{eq:gf-rec-2}
\mathcal{G}_{d+2}(r) &= -\frac{1}{2\pi r} \frac{\partial \mathcal{G}_d(r)}{\partial r} \, .
\end{align}
Using this relation one can rewrite \eqref{solA} in the form
\begin{align}
A_{\alpha}(\ts{x}) = -2\pi\kappa \int \dd^d y \, {j}{}_{\alpha\beta}(\ts{y}) (x^{\beta}-y^{\beta})  \mathcal{G}_{d+2}(\ts{x}-\ts{y}) \, .
\end{align}

For calculations it is convenient to use the following representation of the static Green function given in \cite{Frolov:2015usa},
\begin{align}
\begin{split}
\label{eq:green-function-1}
\mathcal{G}_d(r) &= \frac{1}{(2\pi)^{\tfrac{d}{2}}r^{d-2}} \int\limits_0^\infty \dd\zeta \frac{\zeta^{\tfrac{d-4}{2}}}{a(-\zeta^2/r^2)} J_{\tfrac{d}{2}-1}(\zeta) \, , \\
r^2 &= |\ts{x}-\ts{x'}|^2 \, , \quad d \ge 3 \, .
\end{split}
\end{align}
Last, let us mention that in the limit $r\rightarrow\infty$ the above representation \eqref{eq:green-function-1} gives
\begin{align}
\begin{split}
\label{eq:green-function-local}
\mathcal{G}_d(r) &\sim G_d(r) \quad \text{as } ~ r \rightarrow \infty \, , \\
G_d(r) &= \frac{1}{(2\pi)^{\tfrac{d}{2}}r^{d-2}} \, \lim\limits_{\epsilon\rightarrow 0} \int\limits_0^\infty \dd\zeta \zeta^{\tfrac{d-4}{2}} e^{-\epsilon\zeta} J_{\tfrac{d}{2}-1}(\zeta) \\
&= \frac{\Gamma\left(\tfrac{d}{2}-1\right)}{4\pi^{\tfrac{d}{2}}} \frac{1}{r^{d-2}} \, .
\end{split}
\end{align}
In the above we have made use of the constraint \eqref{eq:constraint}.\footnote{Note that the $\epsilon$-regularization is only required for $d \ge 5$. If one instead calculates the full expression \eqref{eq:green-function-1} for a given choice of the function $a(\Box)$, see a detailed description in Appendix \ref{app:green-functions}, this regularization is not required, and one may simply take the limit $\ell\rightarrow 0$ to recover \eqref{eq:green-function-local}.} $G_d(r)$ is the static Green function of linearized General Relativity \cite{Myers:1986un} in $d$ spatial dimensions, which guarantees that for isolated sources in the far field regime one reproduces the standard asymptotics of General Relativity.

\subsection{Point particles}
The stress-energy of a point-like spinning particle can be written in the form
\begin{align}
{T}{}_{\mu\nu} = \delta{}^t_\mu \delta{}^t_\nu \, {m}\delta{}^{(d)}(\ts{x}) + \delta{}^t_{(\mu} \delta{}^{\alpha}_{\nu)} \, {j}_{\alpha}{}^{\beta} \frac{\partial}{\partial x{}^{\beta}} \delta{}^{(d)}(\ts{x}) \, ,
\end{align}
where $m$ is the mass of the particle and ${j}_{\alpha\beta}$ is a constant antisymmetric matrix parametrizing its angular momentum. A solution for the perturbed metric \eqref{solh}--\eqref{solA} for such a source takes the form\footnote{In four-dimensional spacetime, this solution can be used to obtain a metric for a spinning ring discussed in \cite{Buoninfante:2018xif}.}
\begin{align}
\begin{split}
\label{eq:solution}
{\phi}(r) &= 2\kappa \frac{d-2}{d-1} {m} \, \mathcal{G}_d(r) \, , \\
{A}_{\alpha}(\ts{x}) &= -2\pi\kappa {j}{}_{\alpha\beta} x^{\beta} \, \mathcal{G}_{d+2}(r) \, .
\end{split}
\end{align}
At large distances one recovers the standard expressions known from linearized General Relativity \cite{Myers:1986un}:
\begin{align}
{\phi}(r) &\sim \frac{\Gamma\left(\tfrac{d}{2}\right)}{(d-1)\pi^{\tfrac{d}{2}}} \frac{\kappa m}{r^{d-2}} \, , \\
{A}_{\alpha}(\ts{x}) &\sim -\frac{\Gamma\left(\tfrac{d}{2}\right)}{2\pi^{\tfrac{d}{2}}} \frac{\kappa {j}_{\alpha\beta}x{}^{\beta}}{r^d} \, .
\end{align}
We choose the sign of ${A}_{\alpha}$ such that in the three-dimensional case $d=3$ one obtains the standard Lense--Thirring expression (${j}_{xy} = j$ and $\kappa = 8\pi G$)
\begin{align}
{A}_{\alpha}(\ts{x})\dd x{}^{\alpha} &\sim \frac{2Gj}{r^3} (x\dd y - y \dd x) = \frac{2Gj}{r} \sin^2\theta \, \dd\varphi \, .
\end{align}

\subsection{Extended objects: pencils}
In order to simplify our presentation further, let us consider a special type of spinning objects. That is, we assume that it has finite extension in one spatial directions, while its transverse size is zero. We call such an object a thin spinning pencil or simply ``pencil.''

\subsubsection{Coordinates}
Let us consider two frames. The first one is frame $\bar{S}$ where the matter creating the gravitational field is at rest. The second frame $S$ moves with a constant velocity $\beta$ with respect to $\bar{S}$. We adapt now the choice of the coordinates which is convenient for this situation. Let $\xi$ be a coordinate along the vector of velocity of $S$ and denote by $\ts{x}_{\perp}$ the $d-1$ coordinates orthogonal to the $\xi$-direction. To distinguish the rest frame coordinates from the coordinates in the boosted frame we use a bar for the rest frame coordinates and write
\begin{align}
X^{\mu}=(t,\xi, {x}_{\perp}^i), \quad \bar{X}^{\mu}=(\bar{t},\bar{\xi}, {x}_{\perp}^i)\, .
\end{align}
The index $i=1, 2, \ldots, d-1$ enumerates the coordinates transverse to the direction of motion. We omit the bar for the coordinates $x_\perp^i$ since the Lorentz transformation for the motion in $\xi$-direction does not affect their values. The background Minkowski metric is
\begin{align}
\dd s_0^2 = -\dd \overline{t}^2 + \dd\overline{\xi}^2 + \dd \ts{x}_\perp^2
= -\dd {t}^2 + \dd{\xi}^2 + \dd \ts{x}_\perp^2 \, .
\end{align}
Here, $(\overline{t},\overline{\xi})$ are coordinates in the rest frame $\bar{S}$ and $(t,\xi)$ are the corresponding coordinates in the moving frame $S$. In what follows, we  denote all quantities defined with respect to the rest frame $\bar{S}$  with a bar. For example, the radial distance from the origin to a point $(\bar{\xi},x_{\perp}^i)$ is $\bar{r}^2=\bar{\xi}^2+\ts{x}_{\perp}^2$. Let us specify the $(d-1)$ coordinates $x{}_\perp^j$ orthogonal to the $\bar{\xi}$-direction further:
\begin{align}
\begin{split}
\label{eq:d-epsilion}
x{}_\perp^j &= (y^a, \hat{y}^a, \epsilon z) \, , \quad a = 1, \dots, n \, , \\
n &= \left\lfloor \frac{d-1}{2} \right\rfloor \, , \quad d = 2n+1+\epsilon \, .
\end{split}
\end{align}
One can say that the $(d-1)$-dimensional ``transverse space'' orthogonal to the $\xi$-axis is spanned by $n$ mutually orthogonal two-planes $\Pi_{a}$, and $(y^a, \hat{y}^a)$ are right-handed coordinates in these planes. We shall refer to these planes as \emph{Darboux planes.} If the number of spacetime dimensions $d+1$ is odd one has $\epsilon=1$ and besides these two-planes there exists an additional one-dimensional $z$-axis which is orthogonal to each of the planes as well as to $\xi$-axis. In even spacetime dimensions there is no such additional $z$ coordinate: for example, in four spacetime dimensions there exists only one two-plane orthogonal to the $\xi$-direction. We denote by $\ts{e}^{(a)}=\partial_{y^a}$ and $\hat{\ts{e}}^{(a)}=\partial_{\hat{y}^a}$ unit vectors along  the $y^a$-axis and $\hat{y}^a$-axis, respectively. The 1-forms dual to these vectors are $\ts{\omega}^{(a)}=\dd y^a$ and $\hat{\ts{\omega}}^{(a)}=\dd\hat{y}^a$ such that the volume 2-form for each Darboux plane $\Pi_a$ is given by $\ts{\epsilon}^{(a)}=\ts{\omega}^{(a)}\wedge \hat{\ts{\omega}}^{(a)}$.

\subsubsection{Gravitational field}
The stress-energy tensor of a thin spinning pencil is
\begin{align}
\label{eq:tmunu-pencil}
T_{\mu\nu}=\left[
\delta^{\bar{t}}_{\mu} \delta^{\bar{t}}_{\nu} \bar{\lambda}(\bar{\xi})+\sum\limits_{a=1}^n \left( \bar{j}_a(\bar{\xi}) \delta^{\bar{t}}_{(\mu}
\epsilon^{(a) j} _{\nu)}\partial_j\right)
\right]\delta^{(d-1)}(\ts{x}_{\perp})\, .
\end{align}
We assume that this object has a finite length in $\bar{\xi}$, such that both $\bar{\lambda}(\bar{\xi})$ and $ j_a(\bar{\xi})$ vanish when $\bar{\xi}$ is outside some interval $(0,\bar{L})$. We call $\bar{L}$ the length of the pencil. The mass and the angular momentum of such a pencil are
\begin{align}
\label{eq:m-def}
\bar{m}&=\int \dd\bar{\xi} \, \bar{\lambda}(\bar{\xi})\, ,\\
\label{eq:j-def-1}
\bar{J}_{i j}&=\int \dd\bar{\xi} \, \bar{j}_{ij}(\bar{\xi})\, ,\\
\label{eq:j-def-2}
\bar{j}_{ij}(\bar{\xi})&=\sum\limits_{a=1}^n \epsilon^{(a)}_{ij} \bar{j}_a(\bar{\xi})\, ,
\end{align}
see also Appendix \ref{app:mass-angular-momentum}. The quantities $\bar{\lambda}(\bar{\xi})$ and $\bar{j}_{ij}(\bar{\xi})$ are the mass and angular momentum line densities, respectively. They describe the distribution of the mass and angular momentum along the pencil. In what follows we chose both the total angular momentum $\bar{J}_{i j}$ and its density $\bar{j}_{ij}(\bar{\xi})$ to be orthogonal to the $\bar{\xi}$-direction. Consequently, they have identical Darboux two-planes $\Pi_a$ such that the antisymmetric matrix $\bar{j}_{ij}(\bar{\xi})$ is of the form
\begin{align}
\ts{\bar{j}}&\hat{=} \begin{pmatrix}
0 & \bar{j}_1 & & & \dots & & & 0 \\
-\bar{j}_1 & 0 & & & & & & \\
& & 0 & \bar{j}_2 & & & & & \\
& & -\bar{j}_2 & 0 & & & & \\
\vdots & & & & \ddots & & & \\
& & & & & 0 & \bar{j}_n & \\
& & & & & -\bar{j}_n & 0 & \\
0 & & & & & & & 0
\end{pmatrix} \, .
\end{align}
In the above, the $\bar{j}_a$ are functions of $\bar{\xi}$ alone. By construction, the total angular momentum $\bar{J}_{ij}$ has a similar Darboux form.

The gravitational field $h_{\mu\nu}$ of a thin spinning pencil is
\begin{align}
&\ts{h} = \bar{\phi} \left[\dd t^2+{1\over d-2} (\dd\bar{\xi}^{\,2}+\dd\ts{x}_{\perp}^2)\right]+2 \bar{A}_{i}\dd x_{\perp}^{i}\dd t\, , \label{solh2} \\
&\bar{\phi}(\bar{\xi},x_{\perp}^{i})= 2\kappa \frac{d-2}{d-1} \int \dd\bar{\xi}' \, \bar{\lambda}(\bar{\xi}')\mathcal{G}_d(\bar{r})\, , \label{solPhi2}  \\
 &\bar{A}_{i}(\bar{\xi},x_{\perp}^{i}) =-2\pi \kappa \int \dd\bar{\xi}' \, \bar{j}_{ij}(\bar{\xi}') x_{\perp} ^{ j} \mathcal{G}_{d+2}(\bar{r})\, , \label{solA2}
\end{align}
where we defined the auxiliary expression
\begin{align}
\bar{r}^2 = \left(\bar{\xi} - \bar{\xi}'\right)^2 + \delta{}_{ij} x_\perp^i x_\perp^j \, .
\end{align}
For time-independent objects that are extended also in the transverse direction orthogonal to $\bar{\xi}$ one may use a similar method to construct their gravitational field. Then, however, the energy-momentum \eqref{eq:tmunu-pencil} no longer factorizes in a $\bar{\xi}$-part and a transverse part, but Eq.~\eqref{hhat} still applies.

\section{Ultrarelativistic objects: gyratons}
Now that we have found the gravitational field of a thin pencil in the weak-field limit for any number of dimensions in a wide range of infinite-derivative theories, let us address the ultrarelativistic case arising from performing a boost in the $\overline{\xi}$-direction.

In particular, we shall be interested in the so-called \emph{Penrose limit}. This limit consists of (i) boosting a stationary solution to velocity $\beta$, and then (ii) taking the limit $\beta\rightarrow 1$ while keeping the product $\overline{m}\gamma$ fixed, where
\begin{align}
\gamma = \frac{1}{\sqrt{1-\beta^2}}
\end{align}
is the Lorentz factor, and the mass $\overline{m}$ is given by \eqref{eq:m-def}. Moreover, we shall also assume that $\bar{L}/\gamma$ remains constant during the boost. In this limit the object becomes asymptotically null, and the gravitational fields of these ultrarelativistic objects are called \emph{gyraton fields}.

\subsection{Green function representation}
Before performing the boost, and, subsequently, the Penrose limit, let us briefly mention a useful representation of the static Green function $\mathcal{G}_d(r)$ given by
\begin{align}
\label{eq:green-function-2}
\mathcal{G}_d(r) &= \frac{1}{2\pi}\int\limits_{-\infty}^\infty \frac{\dd\eta}{a(-\eta\ell^2)\eta} \int\limits_{-\infty}^\infty \dd\tau \, K_d(r|\tau) \, e^{i\eta\tau} \, , \\
K_d(r|\tau) &= \frac{1}{(4\pi i \tau)^{\tfrac{d}{2}}} e^{i\tfrac{r^2}{4\tau}} \, .
\end{align}
The function $K_d(r|\tau)$ is the $d$-dimensional heat kernel in imaginary time $\tau = -it$ and therefore satisfies
\begin{align}
\lap K_d(r|\tau) &= -i\partial_\tau K_d(r|\tau) \, , \\
\lim\limits_{\tau\rightarrow 0} K_d(r|\tau) &= \delta{}^{(d)}(\ts{r}) \, .
\end{align}
The derivation of the representation (\ref{eq:green-function-2})  for the static Green function is given in Appendix \ref{app:heat-kernel}. The following property makes this representation very useful for the study of the Penrose limit of solutions: Relation \eqref{eq:green-function-2} expresses the Green function $\mathcal{G}_d(\bar{r})$ as a double Fourier transform, wherein the radius $\bar{r}$ enters only quadratically via the exponential function $\sim\exp[i\bar{r}^2/(4\tau)]$. Since $\bar{r}^2=\bar{\xi}^2+\ts{x}_{\perp}^2$, this exponent can be factorized, which in turn allows one to separate the dependence of the integrand on $\bar{\xi}$ as $\sim \exp[i\bar{\xi}^2/(4\tau)]$. Hence, when applying the boost, only this factor is affected. As we shall demonstrate now, this observation allows us to perform the Penrose limit procedure in a very general and convenient form.

\subsection{Penrose limit}
We now apply the Penrose limit to our previously described linearized potentials of a spinning ``pencil.'' We parametrize the boost in the $\overline{\xi}$-direction via
\begin{align}
\label{eq:boost}
\overline{t} = \gamma\left( t - \beta\xi \right) \, , \quad
\overline{\xi} = \gamma\left( \xi - \beta t \right) \, .
\end{align}
Let us first make a simple remark concerning the scaling properties of the pencil characteristics under a boost transformation \eqref{eq:boost}. We assume that both mass and angular momentum are uniformly distributed along the pencil and their densities in the rest $\bar{S}$ frame, $\bar{\lambda}=\bar{m}/\bar{L}$ and $\bar{j}=\bar{J}/\bar{L}$ are constant. Because of the Lorentz contraction, the length of the same pencil, as measured in the moving frame $S$ is $L=\bar{L}/\gamma$, while its energy is $m=\gamma \bar{m}$. As a result, the linear energy density of the pencil in $S$ frame is $\lambda=\gamma^2 \bar{\lambda}$. In the Penrose limit the energy $m$ is taken to be fixed. Thus the energy density $\lambda$ grows to infinity as $\gamma\to \infty$. To keep it finite, one needs to rescale $\bar{L}\to \gamma \bar{L}$ in the boost process, such that the length $L$ remains unchanged. It is easy to check that under such rescalings the components of the angular momentum remain the same and finite. Note that this is a result of our assumption that the angular momentum density is orthogonal to the direction of motion, because in that case its components are not affected by the boost.

For fixed $\xi$, that is, for a fixed point in frame $S$ one has $\bar{\xi}=-\gamma \beta t + \text{const}$. This means that the frame $S$ moves in the negative direction of $\bar{\xi}$ (``left'') with respect to the rest frame $\bar{S}$. In other words, a pencil which is at rest with respect to $\bar{S}$ moves with a positive velocity in $S$ frame.

\begin{figure}[!hbt]
    \centering
    \vspace{10pt}
    \includegraphics[width=0.48\textwidth]{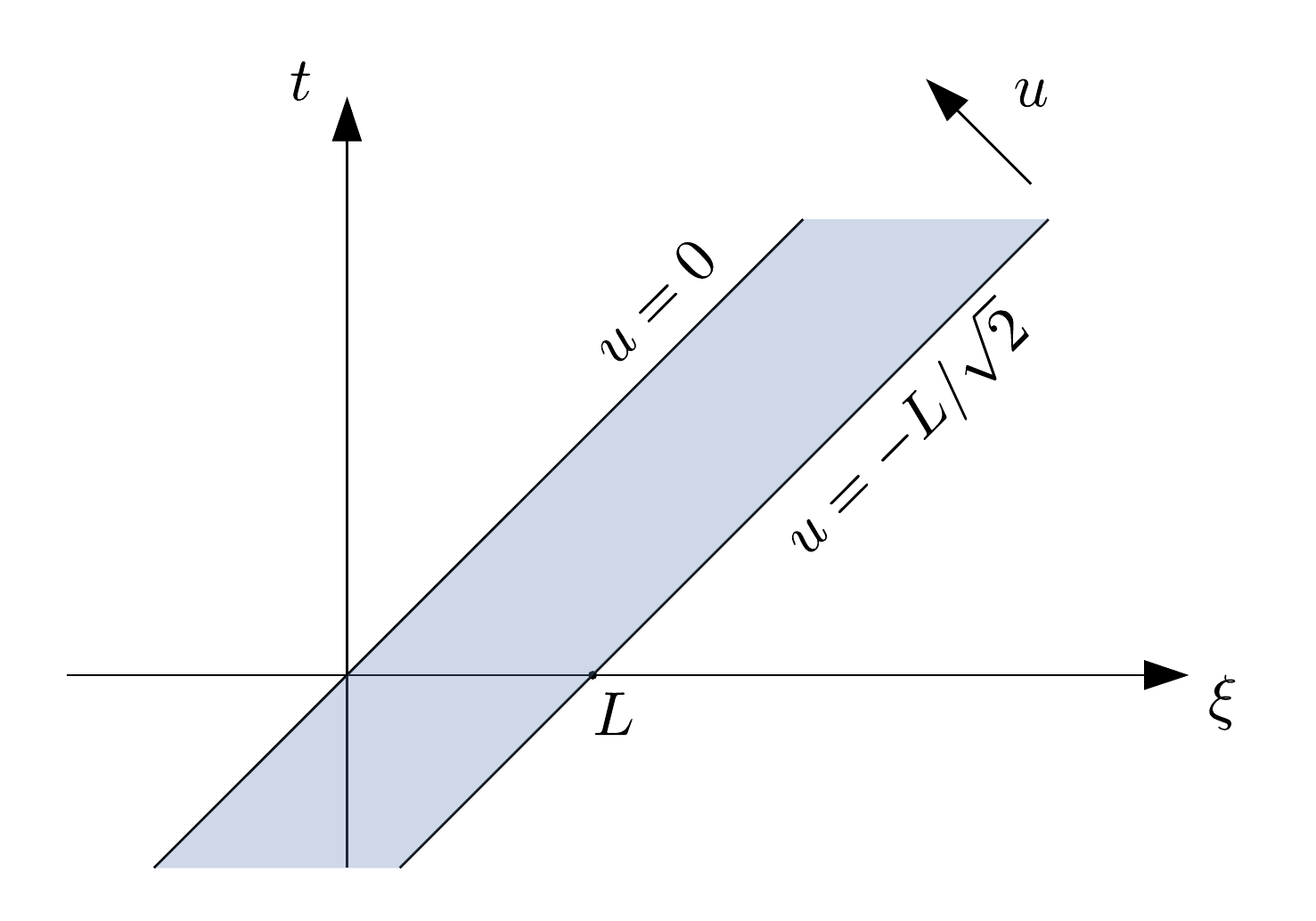}
    \caption{The pencil of length $L$ moves within the two-dimensional $(t,\xi)$-section of Minkowski space in the frame $S$.}
    \label{fig:u-strip}
\end{figure}

We introduce the retarded and advanced null coordinates in the $S$ frame defined as follows:
\begin{align}
u = \frac{t - \xi}{\sqrt{2}} \, , \quad v = \frac{t + \xi}{\sqrt{2}} \, .
\end{align}
Then \eqref{eq:boost} implies
\begin{align}
\label{tx}
\overline{t} &= {\gamma\over \sqrt{2}}[(1+\beta) u+(1-\beta)v]\, ,\\
\overline{\xi} &= {\gamma\over \sqrt{2}}[-(1+\beta) u+(1-\beta)v]\, .
\end{align}
In the ultrarelativistic limit, $\beta \rightarrow 1$, one has
\begin{align}
\overline{t} \to\sqrt{2}\gamma u \, , \quad \overline{\xi} \to -\sqrt{2}\gamma u \, ,
\end{align}
This implies that the matter distribution of such an ultrarelativistec pencil is located in the strip between $u=-L/\sqrt{2}$ and $u=0$ of spacetime; see Fig.~\ref{fig:u-strip}.

Because we keep the ratio $\bar{L}/\gamma$ constant during the Penrose limit, the linear density scales as follows:
\begin{align}
\label{eq:pr-limit-mass}
\lambda(u) = \lim \limits_{\gamma\rightarrow\infty} \sqrt{2} \gamma^2 \, \overline{\lambda}(-\sqrt{2}\gamma u) \, .
\end{align}
This guarantees that in the Penrose limit the product $\overline{m}\gamma$ and the ratio $\bar{L}/\gamma$ remain constant,
\begin{align}
\gamma \, \overline{m} = \gamma \int\limits_{-\infty}^\infty \dd\overline{\xi} \, \overline{\lambda}(\overline{\xi}) = \int\limits_{-\infty}^\infty \dd u \, \lambda(u) = \text{const} \, .
\end{align}
The angular momentum line density $\overline{j}_{ij}(\bar{\xi})$ ``lives'' in transverse space and its tensorial structure is unaffected from the boost. Using this property we define the boosted linear density of the angular momentum in the $S$ frame as follows:
\begin{align}
\label{eq:pr-limit-angular-momentum}
j_{ij}(u)& = \lim\limits_{\gamma\rightarrow\infty} \sqrt{2}\gamma \, \overline{j}_{ij}(-\sqrt{2}\gamma u) \, , \\
j_a(u) &=  \lim\limits_{\gamma\rightarrow\infty} \sqrt{2}\gamma \, \bar{j}_a(-\sqrt{2}\gamma u) \, .
\end{align}
The total angular momentum of the boosted pencil therefore remains finite and has the form
\begin{align}
\begin{split}
J_{ij} &= \int\limits_{-\infty}^\infty\dd\overline{\xi} \, \overline{j}_{ij}(\overline{\xi}) = \int\limits_{-\infty}^\infty \dd u j{}_{ij}(u)\, .
\end{split}
\end{align}

\subsection{Metric}
Under this boost and the Penrose limit, as defined above, the resulting metric takes the form
\begin{align}
\begin{split}
\label{eq:gyraton-1}
\ts{g} &= \left(\eta{}_{\mu\nu} + h{}_{\mu\nu}\right) \dd X{}^\mu \dd X{}^\nu \\
&= -2\dd u \dd v + \phi \dd u^2 + 2 A{}_i\dd x{}^i_\perp \dd u + \dd \ts{x}_\perp^2 \, ,
\end{split}
\end{align}
where we defined
\begin{align}\label{ffAA}
\phi = \lim\limits_{\gamma\rightarrow\infty} 2 \gamma^2 \frac{d-1}{d-2} \bar{\phi} \, , \quad
A_i = \lim\limits_{\gamma\rightarrow\infty} \sqrt{2} \gamma \bar{A}_i \, .
\end{align}
Here, $\bar{\phi}$ and $\bar{A}_i$ are given by \eqref{solPhi2} and \eqref{solA2}, respectively. The integrands in their representations contain the Green function $\mathcal{G}_d(\bar{r})$. In order to understand their behavior under the Penrose limit we make use of relation \eqref{eq:green-function-2}. In this representation the only quantity which is ``sensitive'' to the boost is the heat kernel $K_d$. It factorizes such that the boost-sensitive factor is the exponent of the form $\sim \exp[i(\bar{\xi}-\bar{\xi}')^2/4\tau]$, which for large $\gamma$ factors takes the form $\sim \exp[i\gamma (u-u')^2/2\tau]$. To take the Penrose limit we use the following relation (see also \cite{Shankar:1994}):
\begin{align}
\delta(u) = \lim\limits_{\epsilon\rightarrow 0} \frac{1}{\sqrt{2\pi i\epsilon}}e^{i\tfrac{u^2}{2\epsilon}} \, .
\end{align}
Denote $\epsilon = \tau/\gamma^2$ and apply this relation to \eqref{eq:green-function-2} to obtain 
\begin{align}
\lim\limits_{\gamma\rightarrow\infty} \gamma \, \mathcal{G}_d(\bar{r}) &= \frac{1}{\sqrt{2}} \mathcal{G}_{d-1}(r_\perp) \delta(u-u') \, ,
\end{align}
where $r_\perp^2 = \delta{}_{ij} x^i_\perp x^j_\perp$. Performing the limit $\gamma\to\infty$ in the relations \eqref{ffAA} for the potential $\phi$ and the gravitomagnetic potential $A_i$ finally yields
\begin{align}
\label{eq:phi-final}
\phi &= 2\sqrt{2} \kappa \lambda(u) \mathcal{G}_{d-1}(r_\perp) \,  ,\\
A_i &=-2\pi\kappa j{}_{ij}(u) x^j_\perp \mathcal{G}_{d+1}(r_\perp) \,  .
\end{align}
Introducing polar coordinates $\{\rho_a, \varphi_a\}$ in each Darboux plane $\Pi_a$ such that
\begin{align}
y^a = \rho_a \cos\varphi_a \, , \quad
\hat{y}^a = \rho_a \sin\varphi_a \, ,
\end{align}
one may use the relation
\begin{align}
\label{eq:j_ij_simplified}
j_{ij} \, x_\perp^i \dd x_\perp^j = \sum\limits_{a=1}^n j_a \rho_a^2 \dd\varphi_a
\end{align}
to rewrite the gravitomagnetic potential 1-form as
\begin{align}
\label{eq:a-final}
A_i(\ts{x}_\perp) \dd x{}^i_\perp = 2\pi\kappa \mathcal{G}_{d+1}(r_\perp) \sum\limits_{a=1}^n j_a(u) \rho_a^2 \dd \varphi_a \, ,
\end{align}
which makes the rotational symmetry in each Darboux plane manifest.

\section{Gravitational field of ghost-free gyratons}

In this section we present and discuss gyraton-like solutions  in General Relativity and in infinite-derivative non-local gravity.
In General Relativity, the form factor $a(\Box)$ is simply
\begin{align}
a(\Box) = 1 \,
\end{align}
whereas in infinite-derivative ``ghost-free'' gravity one may postulate instead
\begin{align}
\label{eq:form-factor-gfn}
a(\Box) = \exp\left[(-\Box\ell^2)^N\right] \, .
\end{align}
The static Green function \eqref{eq:green-function-1} can be computed for a wide range of theories, but in the context of the present paper we shall consider General Relativity as well as two infinite-derivative theories corresponding to the choices $N=1$ and $N=2$, which we shall hence refer to as $\mathrm{GF_1}$ and $\mathrm{GF_2}$. It is also possible to extend these studies to arbitrary number of spatial dimensions $d$.

\subsection{Gyratons in $d=3$}

\subsubsection{Gyraton metrics in General Relativity}

As a warm-up, let us consider the well-known gyraton solutions of $(3+1)$-dimensional General Relativity \cite{Bonnor1969,Bonnor:1969rb,Frolov:2005in,Frolov:2005zq}. The relevant two-dimensional and four-dimensional Green functions are
\begin{align}
G_2(r) = -\frac{1}{2\pi} \log(r) \, , \quad
G_4(r) = \frac{1}{4\pi^2r^2} \, .
\end{align}
Since in $d=3$ the transverse space is two-dimensional we have $n=1$ and $\epsilon=0$. Therefore we may write $|\ts{x_\perp}| = \rho$, call the polar angle $\varphi$, and denote by  $j(u)$ the linear density of the angular momentum in the $S$ frame. Then, the gravitational potentials $\phi$ and $\ts{A} = A_i \dd x{}^i$ are
\begin{align}
\phi(u,\rho) &= -\frac{\sqrt{2}\kappa\lambda(u)}{2\pi}\log(\rho) \, , \\
\ts{A}(u) &= \frac{\kappa j(u)}{2\pi} \dd \varphi \, .
\end{align}
This gravitomagnetic field is locally exact such that
\begin{align}
\ts{F} = \dd \ts{A} = 0 \, .
\end{align}
Observe, however, that the gravitomagnetic charge does not vanish:
\begin{align}
Q_0 = \int\limits_\mathcal{A} \ts{F} = \oint\limits_{\partial A} \ts{A} = \kappa j(u) \, .
\end{align}
Here, $\mathcal{A}$ denotes a surface in the Darboux plane. For later convenience we may assume $\mathcal{A}$ to be a circle of radius $\rho$. However, in a given null plane $u=\text{const}$ this charge does not depend on the choice of the contour $\partial \mathcal{A}$. As we shall see soon,  this property is no longer valid in non-local gravity, and effectively the gravitomagnetic current is spread out of the $\rho=0$ line in the direction transverse to the motion.

\subsubsection{Gyraton metrics in ghost-free gravity}
\begin{figure}[!hbt]
    \centering
    \vspace{10pt}
    \includegraphics[width=0.48\textwidth]{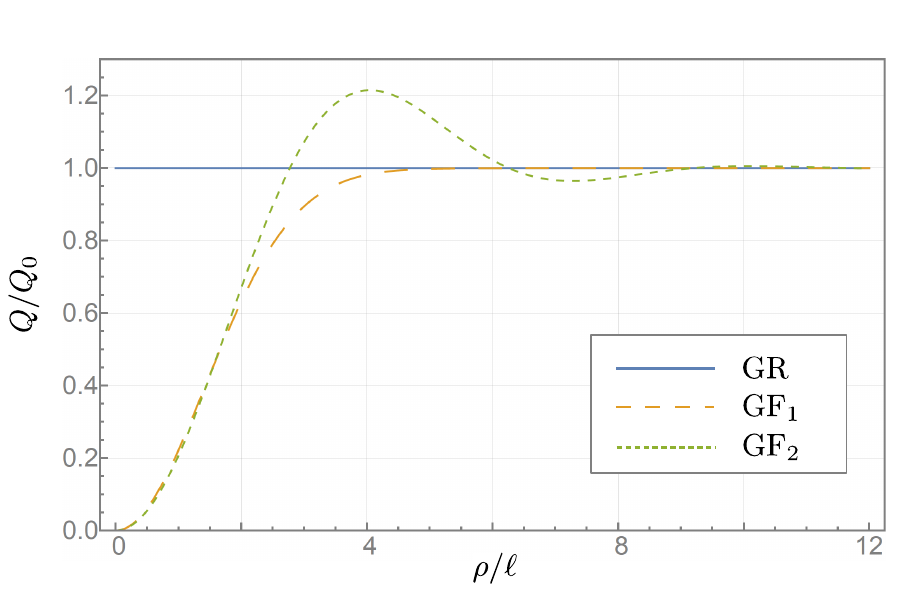}
    \caption{The gravitomagnetic charges on a plane $u=\text{const.}$ of the four-dimensional gyraton in linearized General Relativity as well as linearized $\mathrm{GF_1}$ and $\mathrm{GF_2}$ theory plotted as a function of $\rho/\ell$. The charges are normalized to the value $Q_0$ encountered in General Relativity.}
    \label{fig:charges}
\end{figure}

We consider now a similar gyraton solutions in the non-local theories $\mathrm{GF_1}$ and $\mathrm{GF_2}$. The static Green function for $\mathrm{GF_1}$ theory can be written as
\begin{align}
\mathcal{G}{}_2(r) &= - \frac{1}{4\pi} \text{Ein}\left( \frac{r^2}{4\ell^2} \right) \, ,
\end{align}
where $\text{Ein}(x)$ denotes the complementary exponential integral and $E_1(x)$ is the exponential integral \cite{Olver:2010},
\begin{align}
\text{Ein}(x) &= \int\limits_0^x \dd z \frac{1-e^{-z}}{z} = E_1(x) + \ln x + \gamma \, , \\
E_1(x) &= e^{-x} \int\limits_0^\infty \dd z \frac{e^{-z}}{z+x} = -\text{Ei}(-x) \, ,
\end{align}
and $\gamma = 0.577\dots$ is the Euler--Mascheroni constant. Then, the gravitational potentials $\phi$ and $\ts{A}$ take the form
\begin{align}
\phi(u,\rho) &= -\frac{\sqrt{2}\kappa \lambda(u)}{2\pi}\text{Ein}\left(\frac{\rho^2}{4\ell^2} \right) \, , \\
\ts{A}(u,\ts{x}_\perp) &= \frac{\kappa j(u)}{2\pi}\left[ 1 - \exp\left(-\frac{r_\perp^2}{4\ell^2} \right) \right] \dd \varphi \, .
\end{align}
This gravitomagnetic field is no longer exact and hence the gravitomagnetic charge depends on the radius,
\begin{align}
Q_1(\rho) = \kappa j(u) \left[ 1 - \exp\left(-\frac{\rho^2}{4\ell^2}\right) \right] \, .
\end{align}
At large distances, $\rho \gg \ell$, we recover the gyraton solution obtained in General Relativity. In $\mathrm{GF_2}$ theory one has
\begin{align}
\begin{split}
\mathcal{G}{}_2(r) &= \frac{y}{2\pi} \Big[ \hspace{4pt} \sqrt{\pi}\, {}_1\!F\!{}_3\left(\tfrac12;~1,\tfrac32,\tfrac32;~y^2\right) \\
&\hspace{40pt}- y\, {}_2 \!F\!{}_4\left(1,1;~\tfrac32,\tfrac32,2,2;~ y^2 \right) \Big] \, ,
\end{split}
\end{align}
where we defined $y = \rho^2/(16\ell^2)$. The gravitomagnetic charge now takes the form
\begin{align}
\begin{split}
Q_2(\rho) &= -\kappa j(u) \Big[ 1 - {}_0 F{}_2\left( \tfrac12,\tfrac12; y^2 \right) \\
&\hspace{55pt} - 2\sqrt{\pi}y {}_0 F{}_2 \left( 1, \tfrac32; y^2 \right) \Big] \, .
\end{split}
\end{align}
See Fig.~\ref{fig:charges} for a plot of these charges. Interestingly, the $\mathrm{GF_1}$ charge is monotonic, whereas the $\mathrm{GF_2}$ charge exhibits an oscillatory behavior.

\subsubsection{Curvature invariants}
One may wonder about the geometric properties of the four-dimensional gyraton spacetime
\begin{align}
\begin{split}
\ts{g} &= -2\dd u \dd v + \phi(u,x,y) \dd u^2 + \dd x^2 + \dd y^2 \\
&\hspace{12pt} + 2 \left[ A_x(u,x,y)\dd x + A_y(u,x,y)\dd y \right] \dd u  \, .
\end{split}
\end{align}
This spacetime is a pp-wave because it features a covariantly constant null Killing vector $\ts{k} = \partial_v$ \cite{Stephani:2003tm},
\begin{align}
\nabla_\nu k{}^\mu = 0 \, .
\end{align}
This property remains valid for any choice of the functions $\phi$, $A_x$ and $A_y$, provided their functional dependence remains the same. Since pp-wave spacetimes have vanishing scalar curvature invariants one finds
\begin{align}
R = R{}_{\mu\nu}R{}^{\mu\nu} = R{}_{\mu\nu\rho\sigma}R{}^{\mu\nu\rho\sigma} = 0 \, .
\end{align}
For this reason they remain unchanged for solutions found in the context of linearized infinite-derivative gravity as compared to linearized General Relativity.

\subsection{Gyratons in $d\ge 4$ dimensions}

\subsubsection{$d=4$ case}
In five spacetime dimensions one has $d=4$, which---as per Eq.~\eqref{eq:d-epsilion}---implies that $n=1$ and $\epsilon=1$. In this case there is only one Darboux plane orthogonal to $\xi$ as well as one additional $z$-axis. Let us write the transverse distance as $r_\perp^2 = \rho^2 + z^2$, where $\rho$ is the radial variable in the Darboux plane. Then, from Eqs.~\eqref{eq:phi-final} as well as \eqref{eq:a-final}, one readily obtains
\begin{align}
\phi &= 2\sqrt{2}\kappa\lambda(u)\mathcal{G}_3(r_\perp) \, , \\
A_i \dd x_\perp^i &= -\frac{\kappa}{r_\perp}\frac{\dd}{\dd r_\perp}\mathcal{G}_3(r_\perp) j(u) \rho^2 \dd \varphi \, ,
\end{align}
where $\varphi$ is the polar angle in the Darboux plane, $j(u)$ is the angular momentum eigenfunction, and $\lambda(u)$ describes the density profile. The explicit expressions for the functions $\mathcal{G}_3$ in linearized General Relativity as well as in $\mathrm{GF_1}$ and $\mathrm{GF_2}$ theories are given in Appendix~\ref{app:green-functions}.

\subsubsection{Higher dimensions}

In higher dimensions one can proceed analogously to find expressions for the gyraton metrics. Instead of repeating previous steps, we give here an algorithmic procedure of how to construct such solutions in an arbitrary number of higher dimensions.

First, given the number of spatial dimensions $d$, determine the number of Darboux planes $n$ using \eqref{eq:d-epsilion}. If $d$ is even there will be an independent $z$-axis as well. Due to the rotational symmetry around the pre-boost $\bar{\xi}$-direction it makes sense to introduce polar coordinates in each Darboux plane called $\{\rho_a, \varphi_a\}$ where $a$ labels the Darboux planes. This construction is unique, provided one fixes the direction of the polar angles $\varphi_a$ to be right-handed with respect to the original $\bar{\xi}$-direction.

Second, one introduces the perpendicular radius variable $r_\perp$ according to
\begin{align}
r_\perp^2 = \sum\limits_{a=1}^n \rho_a^2 + \epsilon z^2 \, .
\end{align}
Recall that $\epsilon=1$ if $d$ is even, and $\epsilon=0$ if $d$ is odd. Now one can insert this radius variable into \eqref{eq:phi-final} and \eqref{eq:a-final}. In order to determine the static Green function $\mathcal{G}_d(r_\perp)$ in higher dimensions one may utilize the recursion formulas \eqref{eq:gf-rec-1} and \eqref{eq:gf-rec-2} as well as Appendix \ref{app:green-functions}.

Last, one may want to start with a known line energy density $\bar{\lambda}(\bar{\xi})$ as well as angular momentum line densities $\bar{j}_a(\bar{\xi})$ in the original rest frame. In that case, Eqs.~\eqref{eq:pr-limit-mass} and \eqref{eq:pr-limit-angular-momentum} provide prescriptions as to how to retrieve the resulting functions $\lambda(u)$ and $j_a(u)$ in retarded time.

Realistic gyratons may also have a finite transverse thickness, but due to the linearity of the problem it is always possible to supplement a transverse density function in \eqref{eq:tmunu-pencil} and construct the gravitational field of a ``thick gyraton'' by superposition.

\section{Discussion}
The main goal of this paper is to study the gravitational field of ultrarelativistic spinning objects (gyratons) in the non-local ghost-free theory of gravity. Our starting point is a linearized set of equations for such a theory. In a general case they contain two entire functions of the d'Alembert operator of flat spacetime, $a(\Box)$ and $c(\Box)$, called form factors, subject to the additional constraint that $a(0) = c(0) = 1$. We focused on a simple case when $a(\Box)=c(\Box)$, which guarantees the absence of unphysical modes. The set of field equations in Cartesian coordinates takes the form of uncoupled scalar equations for the components of the gravitational field. When the source is time-independent these equations can be solved by using a static Green function defined as a solution of the equation
\begin{align}
\mathcal{D}\mathcal{G}=-\delta(\ts{x})\hh \mathcal{D}=a(\lap)\lap\, .
\end{align}
In order to obtain a gyraton solution we first found a stationary solution, and then boosted it to the speed of light by means of the Penrose limit. The key observation which allowed us to obtain such a solution is the following: We demonstrated that the Green function $\mathcal{G}$ can be expressed as a double Fourier transform of the heat kernel of the Laplace operator $\lap$. In such a representation all the dependence on the coordinates of the Green function is shifted to the argument of the heat kernel, which has an exponential form $\sim\exp(i\bar{r}^2/4\tau)$, where $\bar{r}$ is the distance between the points which enter as arguments in the Green function, $\bar{r}^2=\bar{\xi}^2+\ts{x}_{\perp}^2$. This exponent can be factorized, such that the dependence on the coordinate $\bar{\xi}$ is universal and has the standard form $\exp(i\bar{\xi}^2/4\tau)$. In the Penrose limit this term produces a delta function of the retarded time $u$, $\delta(u)$. The remaining integral for the Green function $\mathcal{G}_d$ in $d$-dimensional space up to a constant coefficient coincides with $\mathcal{G}_{d-1}$. Thus in the Penrose limit one has schematically
\begin{align}
\mathcal{G}_d\sim \delta(u)\mathcal{G}_{d-1}\, .
\end{align}
It should be emphasized that this result is quite general. In fact, in its derivation we do not use any special form of the operator $\mathcal{D}$, with the exception that $a(z)$ and $c(z)$ are non-zero on the real line.

To obtain explicit gyraton solutions we made additional assumptions. First of all, we chose a form factor $a(\Box)$ of the form \eqref{eq:form-factor-gfn} which guarantees that no extra poles are present in the Green function $\mathcal{G}$. We also assumed the source of the gravitational field to be in the form of an infinitely thin spinning pencil. Then, we applied the boost transformation in the direction of this pencil and chose its internal rotation such that the resulting Darboux two-planes are orthogonal to the direction of motion. In four spacetime dimensions this just means that the pencil is spinning around the axis of the boost direction. Since the field equations in our approximation are linear, one can easily use the found gyraton solutions for infinitely thin pencils to obtain a similar solutions for ``thick'' gyratons, which may have non-trivial structure transverse to the direction of motion.

An interesting but expected property of the obtained ghost-free gyraton metrics is that even for infinitely thin $\delta$-shaped gravitational sources, all solutions are regular at the gyraton axis. This is to be seen in stark contrast to the metrics obtained in linearized General Relativity, wherein the metric functions grow beyond all bounds as $r_\perp\rightarrow 0$, representing a breakdown of the linear approximation scheme. Alternatively, treating the resulting gyraton metrics as geometries beyond the linear approach, the pathological behavior corresponds to a singularity in spacetime. Within linearized ghost-free gravity this pathology disappears entirely, making the linear approximation self-consistent at $r_\perp=0$.

For this reason non-locality effectively spreads the matter and spin distribution of thin gyratons and thereby regularizes them. Another interesting result is that these ghost-free gyraton metrics are vanishing scalar invariant spacetimes: the local curvature invariants vanish. This may be considered as a consequence of a general observation made by Penrose that all metrics after ultrarelativistic boosts take the form of pp-waves \cite{Penrose1976}. In four spacetime dimensions it is known that the Aichelburg--Sexl metric \cite{Aichelburg:1970dh} and its spinning generalisation \cite{Bonnor:1970sb} obtained by boosting linearized solutions of Einstein equation are in fact exact solutions of these non-linear equations. In higher dimensions this is not true \cite{Frolov:2005zq}. However, these higher-dimensional gyraton solutions belong to an important class of so-called Kundt metrics \cite{Stephani:2003tm}. It is interesting to check whether this is also true in complete (non-linear) ghost-free gravity.

Let us finally mention that the ghost-free gyraton solutions obtained in this paper may be used to study the collision of ultrarelativistic particles. In particular, they will allow one to understand the role of non-locality and spin in the process of micro-black hole formation \cite{Yoshino:2007ph,Frolov:2015bta}.

\section*{Note Added in Proof}

It has been brought to our attention that exact pp-wave solutions in ghost-free infinite-derivative gravity have been studied in Ref.~\cite{Kilicarslan:2019njc}. Recently, these studies have been extended to exact wavelike ``impulsive'' solutions in anti-de Sitter spacetimes \cite{Dengiz:2020xbu}, and it would be very interesting to understand the role of the gyraton metrics obtained in this paper in that context.

\section*{Acknowledgments}
J.B.\ is grateful for a Vanier Canada Graduate Scholarship administered by the Natural Sciences and Engineering Research Council of Canada as well as for the Golden Bell Jar Graduate Scholarship in Physics by the University of Alberta. V.F.\ thanks the Natural Sciences and Engineering Research Council of Canada and the Killam Trust for their financial support.

\appendix

\section{Mass and angular momentum of extended objects in higher dimensions}
\label{app:mass-angular-momentum}

We denote by $X^{\mu}=(t,x^{\alpha})$ Cartesian coordinates in $d+1$ dimensional Minkowski spacetime and use indices $\alpha, \beta, \ldots =1,2,\ldots ,d$ from the beginning of the Greek alphabet to label spatial coordinates. Let us consider distribution of matter described by the stress-energy of the form
\be\n{a1}
T_{00}=\rho(\ts{x}), \quad T_{0\alpha}=\frac12{\partial \over \partial x^ \beta} j_{\alpha  \beta}(\ts{x})\, ,\quad T_{\alpha  \beta}=0\, .
\ee
where $j_{\alpha  \beta}(\ts{x})$ is an anti-symmetric tensor function. It is easy to check that this stress-energy tensor satisfies the required conservation law $\partial{}_\mu T^{\mu\nu}=0$. Denote by $\ts{\xi}_{(\mu)}$ a generator of the space-time translations, and by $\ts{\zeta}_{(\alpha  \beta)}$ the generators of the rigid spatial rotations, then one has
\begin{align}
\label{a2}
\ts{\xi}_{(\mu)}&=\xi_{(\mu)}^{\nu}\partial_{\nu}=\partial_{\mu}\, ,\\
\ts{\zeta}_{(\alpha\beta)}&=\ts{\zeta}^\nu_{(\alpha\beta)}\partial{}_\nu = x_\alpha\partial_{ \beta} -x_ \beta\partial_{\alpha}\, .
\end{align}
The conserved quantities related to these symmetries are
\ba\n{cons}
P_{\mu}&=&\int \dd^d x \, T_{0\nu}\xi_{(\mu)}^{\nu}\, ,\\
J_{\alpha  \beta}&=&\int \dd^d x \, T_{0 \gamma}\ts{\zeta}_{(\alpha  \beta)}^{\gamma}\, .
\ea
or in an explicit form
\ba
M&=&P_{0}=\int \dd^d x \, T_{00}\, , \n{mpj1}\\
P_{\alpha}&=&\int \dd^d x \, T_{0\alpha}\, ,\n{mpj2}\\
J_{\alpha  \beta}&=&\int \dd^d x \, (x_\alpha T_{0 \beta}- x_ \beta T_{0\alpha})\, .\n{mpj3}
\ea
We assume that the stress-energy tensor (\ref{a1}) either vanishes outside some compact region, or it is sufficiently fast decreasing at far spatial distance, so that the surface terms arising as a result of integration by parts in (\ref{mpj3}) vanish. Simple calculations give
\be
M=\int \dd^d x \, T_{00} \hhh P_{\alpha}=0\hhh J_{\alpha  \beta}=\int \dd^d x \, j_{\alpha  \beta}\,  .
\ee
The relation $P_{\alpha}=0$ implies that the stress-energy tensor (\ref{a1}) is written in the center of mass frame.

\begin{widetext}
\section{Heat kernel representation of ghost-free static Green functions}

\label{app:heat-kernel}
The static Green function $\mathcal{G}_d$ considered in this paper satisfies the relation
\begin{align}
\label{eq:app:gf-def}
a(\lap)\lap\mathcal{G}_d(\ts{x},\ts{x'}) = -\delta(\ts{x}-\ts{x'})
\end{align}
Here $\lap$ is a Laplace operator in $d$-dimensional space.  We denote by $K_d(\ts{x}|\tau)$ the $d$-dimensional heat kernel of $\lap$. It is defined as a solution of the equation
\be 
 \lap K_d(\ts{x}|\tau) = -i\partial_\tau K_d(\ts{x}|\tau) \, ,
\ee
obeying the boundary conditions
\begin{align}
\label{eq:app:k-bdy}
\lim\limits_{\tau\rightarrow 0} K_d(\ts{x}|\tau) = \delta(\ts{x}) \, , \quad \lim\limits_{\tau\rightarrow \pm\infty} K_d(\ts{x}|\tau) = 0 \, .
\end{align}
It has the following explicit form:
\be
\label{eq:app:h-properties}
K_d(\ts{x}|\tau) = \frac{1}{(4\pi i \tau)^{d/2}} \exp\left( \frac{i \ts{x}^2}{4\tau} \right) \,   .
\ee
Let us define the object $\mathcal{K}_d(\ts{x}|\tau)$ as a solution of the equation
\begin{align}
\label{eq:app:ktilde-def}
a(\lap)\mathcal{K}_d(\ts{x}|\tau) = i K_d(\ts{x}|\tau) \, .
\end{align}
Then it is easy to check the required Green function $\mathcal{G}_d$ can be written in the form
\begin{align}
\label{eq:app:gf-rep}
\mathcal{G}_d(\ts{x},\ts{x'}) = \int\limits_0^\infty \dd\tau \, \mathcal{K}_d(\ts{x}-\ts{x'}|\tau) \, .
\end{align}
We introduce now the Fourier transform of $\mathcal{K}_d$ and its inverse by means of the relations
\begin{align}
\widetilde{\mathcal{K}}_d(\ts{x}|\omega) = \int\limits_{-\infty}^\infty \dd\tau \, e^{i\omega\tau} \mathcal{K}_d(\ts{x}|\tau) \, , \quad
\mathcal{K}_d(\ts{x}|\tau) = \int\limits_{-\infty}^\infty \frac{\dd\omega}{2\pi} \, e^{-i\omega\tau} \widetilde{\mathcal{K}}_d(\ts{x}|\omega) \, .
\end{align}
Then we may write
\begin{align}
\mathcal{G}_d(\ts{x},\ts{x'}) &= \int\limits_0^\infty \dd\tau \int\limits_{-\infty}^\infty \frac{\dd\omega}{2\pi} \int\limits_{-\infty}^\infty \dd\tau' e^{-i\omega(\tau-\tau')} \mathcal{K}_d(\ts{x}-\ts{x'}|\tau') \\
&= \int\limits_0^\infty \dd\tau \int\limits_{-\infty}^\infty \frac{\dd\omega}{2\pi} \int\limits_{-\infty}^\infty \dd\tau' e^{-i\omega(\tau-\tau')} \frac{i}{a(\lap)}K_d(\ts{x}-\ts{x'}|\tau') \\
&= \int\limits_0^\infty \dd\tau \int\limits_{-\infty}^\infty \frac{\dd\omega}{2\pi} \int\limits_{-\infty}^\infty \dd\tau' e^{-i\omega(\tau-\tau')} \frac{i}{a(-i\partial_\tau)}K_d(\ts{x}-\ts{x'}|\tau') \\
&= \int\limits_0^\infty \dd\tau \int\limits_{-\infty}^\infty \frac{\dd\omega}{2\pi} \int\limits_{-\infty}^\infty \dd\tau' e^{-i\omega(\tau-\tau')} \frac{i}{a(-\omega)}K_d(\ts{x}-\ts{x'}|\tau') \, .
\end{align}
In the first equality we have used \eqref{eq:app:ktilde-def}, then used the properties of the heat kernel via Eq.~\eqref{eq:app:h-properties}, and finally integrated by parts where the boundary terms vanish due to \eqref{eq:app:k-bdy}. The integral over $\tau$ can be easily calculated assuming that one takes care about its asymptotic behavior and uses the standard regularization. By using the relation
\begin{align}
\int\limits_0^\infty \dd\tau e^{-i\omega\tau} \equiv \lim\limits_{\epsilon\rightarrow 0}\int\limits_0^\infty \dd\tau e^{-i(\omega-i\epsilon)\tau} = \frac{-i}{\omega} \, .
\end{align}
one obtains
\begin{align}
\mathcal{G}_d(\ts{x},\ts{x'}) &= \int\limits_{-\infty}^\infty \frac{\dd\omega}{2\pi} \int\limits_{-\infty}^\infty \dd\tau' e^{i\omega\tau'} \frac{1}{\omega a(-\omega)}K_d(\ts{x}-\ts{x'}|\tau')\, ,
\end{align}
which is the double Fourier representation for the Green function $\mathcal{G}_d$ used in the main body of the paper.

\section{Static infinite-derivative ghost-free Green functions}
\label{app:green-functions}
Let us consider theories with the form factor $a(\lap)$ of the form $a{}^N(\lap) = \exp\left[(-\lap\ell^2)^N\right]$, where $N$ is a positive integer number. We refer to such a theory as ghost-free gravity and use the abbreviation $\mathrm{GF_N}$ for such a theory. For $N=0$, $a^0(\lap)=1$ and the corresponding theory is nothing but linearized General Relativity. Let us write $\mathcal{D}_N=a{}^N(\lap)\lap$ and denote by $\mathcal{G}_d^N$ a static Green function for $\mathrm{GF_N}$ theory in a space with $d$ dimensions. Such a Green function obeys the equation
\begin{align}
\mathcal{D}_N \mathcal{G}_d^N (r)=-\delta^{(d)} (\ts{r})\, .
\end{align}
For $N=0$, that is, in General Relativity, we also use the notation $G_d (r)=\mathcal{G}_d^0 (r)$. The static Green functions can be found by using Eqs.~\eqref{eq:gf-rec-2}--\eqref{eq:green-function-1}. In this appendix we collect exact expressions for these Green functions for General Relativity as well as $\mathrm{GF_1}$ and $\mathrm{GF_2}$ theory for the number of spatial dimensions $d=1,2,3,4$. Using the recursive relations \eqref{eq:gf-rec-2} one can obtain their expression for $d\ge 5$. In what follows we will use the abbreviation $y =( r/4\ell)^2 $.
\begin{align}
G{}_1(r) &= -\frac{r}{2} \, , \\
\mathcal{G}_1^1(r) &= -\frac r2 \text{erf}\left(\frac{r}{2\ell}\right) - \ell\frac{\exp{\left[-r^2/(4\ell^2)\right]} - 1}{\sqrt{\pi}} \, , \\
\mathcal{G}_1^2(r) &= -\frac{\ell}{\pi} \Big\{ \hspace{6pt} 2\Gamma(\tfrac14) y\, {}_1\!F\!{}_3\left( \tfrac14;~ \tfrac34,\tfrac54,\tfrac32;~ y^2 \right) + \Gamma(\tfrac34) \Big[ {}_1\!F\!{}_3\left( -\tfrac14;~ \tfrac14,\tfrac12,\tfrac34;~ y^2 \right) - 1 \Big] \Big\} \, \\
G{}_2(r) &= -\frac{1}{2\pi}\log\left(\frac{r}{r_0}\right) \, , \\
\mathcal{G}_2^1(r) &= - \frac{1}{4\pi} \text{Ein}\left( \frac{r^2}{4\ell^2} \right) \, , \\
\mathcal{G}_2^2(r) &= -\frac{y}{2\pi} \Big[ \hspace{4pt} \sqrt{\pi}\, {}_1\!F\!{}_3\left(\tfrac12;~1,\tfrac32,\tfrac32;~y^2\right) - y\, {}_2 \!F\!{}_4\left(1,1;~\tfrac32,\tfrac32,2,2;~ y^2 \right) \Big] \, , \\
G_3(r) &= \frac{1}{4\pi r} \, , \\
\mathcal{G}_3^1(r) &= \frac{\text{erf}[r/(2\ell)]}{4\pi r} \, , \\
\mathcal{G}_3^2(r) &= \frac{1}{6\pi^2\ell}\Big[ 3 \Gamma\!\left(\tfrac54\right) {}_1\!F\!{}_3\left( \tfrac14;~ \tfrac12,\tfrac34,\tfrac54;~ y^2 \right) - 2y\Gamma\!\left(\tfrac34\right) {}_1\!F\!{}_3\left( \tfrac34;~ \tfrac54, \tfrac32, \tfrac74;~ y^2 \right) \Big] \, , \\
G_4(r) &= \frac{1}{4\pi^2 r^2} \, , \\
\mathcal{G}_4^1(r) &= \frac{1 - \exp\left[-r^2/(4\ell^2)\right]}{4\pi^2 r^2} \, , \\
\mathcal{G}_4^2(r) &= \frac{1}{64\pi^2 y\ell^2}\Big[ 1 - {}_0\!F\!{}_2\left( \tfrac12,\tfrac12;~ y^2 \right) + 2\sqrt{\pi} y\, {}_0\!F\!{}_2\left( 1, \tfrac32;~ y^2 \right) \Big] \, .
\end{align}
Here we use the standard notation ${}_a F_{b}$ for the hypergeometric function \cite{Olver:2010}.
\end{widetext}

\bibliography{Ghost_references,GYRATON}{}

\end{document}